\documentclass[12pt]{article}
\pdfoutput=1
\usepackage{jhep-mod2}
\usepackage{bm}
\usepackage{amssymb}% http://ctan.org/pkg/amssymb
\usepackage{pifont}% http://ctan.org/pkg/pifont
\usepackage{slashed} % for Dirac operator
\usepackage{slantsc} % slanted smallcaps...
\begin{document}
%-------------------------------------------------------------------------------------------
\title{The utterly prosaic connection between physics and mathematics}
%-------------------------------------------------------------------------------------------
\author{Matt Visser}
\affiliation{School of Mathematics and Statistics, \\
Victoria University of Wellington, PO Box 600, Wellington 6140, New Zealand}
\emailAdd{matt.visser@sms.vuw.ac.nz}
%-------------------------------------------------------------------------------------------
\abstract{

\noindent
Eugene Wigner famously argued for the ``unreasonable effectiveness of mathematics''  for describing physics and other natural sciences in his 1960 essay. That essay has now led to some 55 years of (sometimes anguished) soul searching --- responses range from ``So what? Why do you think we developed mathematics in the first place?'', through to  extremely speculative ruminations on the existence of the universe (multiverse) as a purely mathematical entity --- the Mathematical Universe Hypothesis. In the current essay I will steer an utterly prosaic middle course: Much of the mathematics we develop is informed by physics questions we are tying to solve; and those physics questions for which the most utilitarian mathematics has successfully been developed are typically those where the best physics progress has been made. 

\bigskip
\noindent
February 2015; March 2017; \LaTeX-ed \today

\bigskip
\noindent
{Essay written for the 2015 FQXi essay contest:\\
\centerline{ ``Trick or truth: The mysterious connection between physics and mathematics".}
}
}
\notoc
 
%-------------------------------------------------------------------------------------------
%-------------------------------------------------------------------------------------------
%-------------------------------------------------------------------------------------------
\maketitle
%-------------------------------------------------------------------------------------------
\def\d{{\mathrm{d}}}
%-------------------------------------------------------------------------------------------
\clearpage
%-------------------------------------------------------------------------------------------

%-------------------------------------------------------------------------------------------
\section{Background} 
%-------------------------------------------------------------------------------------------
%-------------------------------------------------------------------------------------------
%-------------------------------------------------------------------------------------------
\parskip5pt
%-------------------------------------------------------------------------------------------

Mathematics is simply a way of codifying, in an abstract manner, various regularities we observe in the physical universe around us.   
Many trees have been sacrificed on the altar of obfuscation, in an attempt to make the situation appear more excessively mystical than it ultimately  is.  The problem is not with mysticism \emph{per se}, but with excessive mysticism used as a tool to obfuscate --- and to disguise limited competency and charlatanism or worse.

I shall argue that  the apparent ``unreasonable effectiveness'' of mathematics in the natural sciences~\cite{Eugene} is largely an illusion; there is a quite natural back-and-forth between mathematics and the natural sciences --- a dialectic --- whereby some branches of mathematics are preferentially worked on because they are so useful for the natural sciences, and some branches of the natural sciences make great leaps in understanding because the related mathematics is so well developed.  This does not, however, mean that progress occurs in lock-step --- sometimes mathematical formalism out-runs what the natural philosophers can measure/observe; sometimes the experimental/observational abilities of the natural philosophers out-run what the mathematicians can usefully analyze. Sometimes, (even in the modern world), progress can be out of phase by decades, (more rarely,  even by centuries).

One of the great success stories in this back-and-forth has been the development and flowering, in the decades and centuries since 1666, of the differential and integral calculus; this was largely (and with due acknowledgement to the sometimes fractious and tendentious behaviour of some of the personalities involved) a collaborative effort between the natural sciences and the mathematical sciences. While it took decades to develop useful computational  tools, it took almost two centuries to make everything rigorous. 
One can still argue over the precise experimental/observational relationship between infinitesimal calculus and empirical reality~\cite{best}, but there is absolutely no doubt that infinitesimal calculus is pragmatically extremely useful over an extraordinarily wide range of situations. The \emph{continuum} of differential calculus is, and for the foreseeable future is likely to continue to be, a key component in our understanding of physical aspects of empirical reality. 

\enlargethispage{15pt}
In contrast the \emph{discretium} of the discrete mathematicians has had lesser direct impact in physics itself, instead having more direct applications in computer science, (particularly, algorithm analysis), operations research,  and to some extent in the social sciences. 
Again, in those fields, there is considerable back-and-forth between practitioners in those fields, and mathematicians either developing the relevant mathematics, or playing catch-up by codifying, making rigorous, and justifying empirically derived heuristics. 

\clearpage

%-------------------------------------------------------------------------------------------
\section{Quantum conundrums} 
%-------------------------------------------------------------------------------------------

Quantum physics is another scientific enterprise where there has been much back-and-forth between the mathematics and physics communities. Some of the physicists founding the field of quantum physics had very pronounced mystical leanings~\cite{Q1,Q2}, which in the hands of true experts is not necessarily a bad thing. In the hands of lesser mortals however, mysticism can often lead to unnecessary and excessive obfuscation~\cite{Q3}. 

%-------------------------------------------------------------------------------------------
\subsection{Quantum pedagogy} 
%-------------------------------------------------------------------------------------------

The notion of ``quantum complementarity'' is often phrased in terms of a quantum object being \emph{either} a wave \emph{or} a particle, but \emph{never both} at the same time. \emph{This} specific notion of quantum complementarity is seriously defective --- and in my opinion close to vacuous --- it is less than clear what is actually being asserted. A more nuanced version of quantum complementarity is the more precise and sensible physics statement that whenever two quantum observables (Hermitian operators) do not commute then approximate knowledge of one observable constrains the extent to which one can determine the other --- but I have no idea how to convert the phrases ``this quantum object is a particle" and ``this quantum object is a wave'' into Hermitian operators on Hilbert space.
This more nuanced version of quantum complementarity is effectively a variant of the Heisenberg uncertainty relation. 

But the Heisenberg uncertainty relations (as commonly presented) suffer from their own level of excessive mysticism and obfuscation.
A distressingly common (but utterly infantile) confusion is to treat the words ``quantum'' and ``non-commutative'' as though they are synonyms; they are not. There are many non-commutative objects in both mathematics and physics that simply have nothing to do with quantum effects; the two concepts overlap, but are by no means identical. 

\enlargethispage{10pt}
Perhaps worse, there is also a distressingly common misconception that ``uncertainty relations'' are intrinsically a quantum phenomenon --- utterly ignoring the fact that engineers have by now some 60 years of experience with utterly classical time-frequency uncertainty relations in signal processing, and that mathematicians have by now over 80 years experience with utterly classical time-frequency uncertainty relations in Fourier transform theory. The ``central mystery'' in quantum physics is not the Heisenberg uncertainty relations --- the central mystery  is instead de Broglie's momentum-wavenumber relation $p = \hbar k$, and Einstein's energy-frequency relation $E=\hbar\omega$. It is these relations, which inter-twine the particle aspects of the quantum object with the wave aspects, and so  lead to the concept ``wavicle'', that are utterly central to the quantum enterprise. 

Finally, let me mention ``tunnelling''/``barrier penetration''. Despite yet more common misconceptions, tunnelling is a simply wave phenomenon; it is not (intrinsically) a quantum physics phenomenon.  Under the cognomen ``frustrated total internal reflection'', the classical tunnelling phenomenon has been studied and investigated for well over 300 years, with the wave aspects (the ``evanescent wave'') coming to the foreground  approximately 150 years ago, well before the formulation \emph{circa} 1900 of even the most basic of quantum concepts (Planck's blackbody radiation spectrum).

Now the examples I have been discussing in this last page are, properly speaking, neither problems of physics nor problems of mathematics --- they are problems of pedagogy and presentation. They should serve to remind us that some thought should be put into communicating clearly; not necessarily precisely. (Excessive precision is the hobgoblin of small minds; as anyone who has ever taught freshman calculus or freshman physics can attest --- clarity is typically more important than precision.)

%-------------------------------------------------------------------------------------------
\subsection{Quantum foundations} 
%-------------------------------------------------------------------------------------------

So where are the ``real'' open problems in quantum physics? There are certainly many technical problems in (relativistic) quantum field theory (see below), but the truly foundational open issues have to do with the so-called ``measurement problem'' and the ``collapse of the wavefunction''; issues that continue to plague quantum physics even after some 90 years. This is a physics problem, not a mathematics problem, and almost certainly will not need ``new mathematics'' for its resolution. 
Despite multiple and very loud claims to the contrary, quantum decoherence is simply not enough. 

\enlargethispage{10pt}
At best, quantum decoherence might reduce quantum amplitudes to classical probabilities --- but this is still missing the last essential step --- ``reification" the ``making real" of \emph{one} specific outcome, \emph{one} specific unit of history. Quantum physics in its current state simply cannot explain history --- the observed fact that (as far as we can tell) the universe really does have a single unique past history. If one takes the usual Feynman ``sum over histories" seriously, then (without some \emph{realist} solution to this problem) the notion of a single past history simply does not exist. While I am sure that there are many political operators (from all over the political spectrum) who would like to use (abuse) quantum physics to undermine the notion of history, (and so undermine historical responsibility for past actions), at some stage one simply has to pay attention to observed reality and the historical record. Normally the ``measurement problem"  is phrased in terms of outcomes for future experiments; but philosophically, (and also in terms of the underlying physics), it is the past that is the crucial issue. 

\clearpage

Without some real  physical mechanism for the ``collapse of the wavefunction", there simply is no physical basis for the notions of memory, or history,  a circumstance which then utterly undermines (even as an approximation) any notion of classical physics and the very notion of causality. To have a notion of definite history one needs, at the very least, a dense network of classical collapse events, densely spaced in both space and time --- at least in our past causal cone.
Quantum physics would then live ``in the gaps" between the collapse events, and the ``collapse of the wavefunction" would have to be an objective feature of physical reality. 
(Probably the best-known models of this type are the  Ghirardi--Rimini--Weber [GRW] model and the Penrose model; both of which suffer from quite serious physics  limitations; but more importantly have suffered from malicious neglect by the physics community.) 
Despite the unquestioned success of the ``shut up and calculate'' non-interpretation of quantum physics, there are real physics issues still to be dealt with in the foundations of the subject. Consider for instance the  \emph{fictional} musings of  the \emph{fictional} physicist Shevek~\cite{Le-Guin}:
\begin{quote}
...the physicists of [Einstein's] own world had turned away from
his effort and its failure, pursuing the magnificent incoherencies of
quantum theory, with its high technological yields, ... to arrive at a
dead end, a catastrophic failure of the imagination.\\
\null\hfill--- Shevek \emph{circa} 2500 CE
\end{quote}

%-------------------------------------------------------------------------------------------
\subsection{Quantum field theory} 
%-------------------------------------------------------------------------------------------

In contrast to the foundational physics problems considered above, the purely technical problems facing quantum field theory are more fundamentally mathematical in nature. While the mathematical aspects of quantum mechanics were placed on a firm foundation by von Neumann and others, the mathematical foundations of quantum field theory are much shakier. Consider for instance the well-known comments:
\begin{quote}
In the thirties, under the demoralizing influence of 
quantum-theoretic perturbation theory, the mathematics
required of a theoretical physicist was reduced to a 
rudimentary knowledge of the Latin and Greek alphabets.
\\
\null\hfill--- Res Jost \emph{circa} 1964
\end{quote}
\begin{quote}
I am acutely aware of the fact that the marriage 
between mathematics and physics, which was so 
enormously fruitful in past centuries, has recently 
ended in divorce.
\\
\null\hfill--- Freeman Dyson 1972
\end{quote}
What is going on  here? While no one doubts that the quantum field theory representing the standard model of particle physics is a great success --- as physics --- the more mathematically inclined members of the physics community are less than happy with technical aspects of the situation... There are a number of issues:
\begin{itemize}
\item 
The fact that the Feynman expansion cannot possibly converge, (it is at best an asymptotic expansion even after you renormalize to effectively make each individual Feynman diagram finite), is probably just an annoyance...
\item
As of December 2014 Haag's theorem was still an obstruction to constructing a fully relativistic interaction picture, rather completely undermining standard textbook presentations of how to derive the Feynman diagram expansion. This \emph{may} have been fixed (or rather, side-stepped) as of January 2015~\cite{Haag}.
\item
As of February 2015,  not one single non-trivial interacting relativistic quantum field theory has rigorously been established to exist in 3+1 dimensions; though rigorous constructions are available in 2+1 and 1+1 dimensions. (The technical difference seems to be that there are interesting super-renormalizable quantum field theories in (2+1) and (1+1) dimensions; but that the rigorous techniques used to establish these results to not quite extent to the physically interesting, but merely renormalizable, quantum field theories in 3+1 dimensions.)
\end{itemize}
Almost certainly these are merely nasty technical annoyances; places where the mathematics has not yet caught up with the physics. 
Remember that after Newton and Leibniz it took almost 200 years before calculus was put on a really firm mathematical foundation. This did not stop physicists and others from \emph{using} calculus during that two century interregnum, and using it to good purpose and effectiveness. Similarly, the  standard model of particle physics clearly has very many features that are undoubtedly correct, very many features that are undoubtedly good representations of empirical reality, so physicists will continue to use it regardless of what the mathematicians feel about the technical details.

%-------------------------------------------------------------------------------------------
\section{Usability versus precision} 
%-------------------------------------------------------------------------------------------

The key issue here is usability versus precision; while precision is sometimes important usability will always trump precision. 
This is a variant of the old debate between accuracy and precision; there are sometimes cases when precision (obtaining many decimal places) is important, but saner people will prefer accuracy (fewer decimal places; but ones that are actually correct). 

This observation is important because, as long as there is plenty of reliable experimental/observational data coming in, then empirical reality has a way of trumping sloppy theory. 
Theorists can often afford to cut a few corners, (and be, from a mathematical perspective, more than a little bit sloppy), as long as there is steady stream of data to keep them on track.   
If the data-steam dries up, considerably more care is called for --- the techniques that are effective in a data-rich environment, can quite easily and unfortunately lead to ``forty years wandering in the wilderness'' in a data-poor environment. This is partly the reason pure mathematicians make such a fetish of precision --- since they are typically (not always) working at a greater remove from empirical data. 

Interruptions in the data-stream can come from at least two sources --- possible technological limitations and/or sociological issues. Sometimes we just cannot collect the data because the equipment to do so simply does not exist; sometimes the equipment exists but collecting the data would be grossly unethical. 
Sometimes the barriers are more subtle --- even in Western societies over the last century and a half there have been occasions when experimentalists have looked down upon theorists and vice versa; sometimes tribalism \emph{within} the theoretical physics community has hindered progress. (Remember the phrase ``squalid state physics''? I have heard considerably worse.) Sometimes communication between mathematicians and physicists has essentially ground to a halt.

So the close connection between mathematics and physics is dynamic not static; the back-and-forth connections between the two will continually twist and strain in response to technological limitations and the personalities involved. After all, both mathematics and physics are in the end human endeavours; and human beings are a perhaps excessively refractory material to deal with.

%-------------------------------------------------------------------------------------------
\section{Mathematical universe hypothesis} 
%-------------------------------------------------------------------------------------------

\begin{quote}
God created the integers; all else is the work of man.\\
\null\hfill---Leopold Kronecker
\end{quote}

\begin{quote}
Don't let me catch anyone talking about the universe in my department.\\
\null\hfill---Ernest Rutherford
\end{quote}

At its most extreme the undoubtedly close connection between mathematics and physics is sometimes asserted to be an \emph{identity} --- this is the ``mathematical universe hypothesis''~\cite{Tegmark1, Tegmark2}. 

I personally think this is excessive, unnecessary, and simply not useful.  We do not need to imbue mathematics with more significance than it already undeniably has --- the abstract codification of regularities in the empirical universe is quite enough. 
Ironically the ``shut up and calculate'' point of view advocated in~\cite{Tegmark2} does not actually imply the mathematical universe hypothesis. 

``Shut up and calculate''  is a non-interpretation, a non-ontology which requires nothing specific in the way of a philosophical commitment; similarly the ``shut up and calculate'' non-interpretation of quantum physics 
requires nothing specific in the way of a philosophical commitment.   Most physicists would agree with the ``external universe hypothesis'', but the gap between the ``external universe hypothesis''  and the  ``mathematical universe hypothesis'' is a very large one, and the logic connecting the two is not at all convincing. 

For instance, it is asserted~\cite{Tegmark1} that  ``The [mathematical universe hypothesis] makes the testable prediction that further mathematical regularities remain to be uncovered in nature.'' This is not exactly a unique distinguishing characteristic of the mathematical universe hypothesis --- just about any random philosophy of physics would predict that ``further mathematical regularities remain to be uncovered in nature''; even in the ``shut up and calculate'' non-interpretation one would hardly be surprised if further mathematical regularities were to be found. 

More alarmingly, the level I to level IV mathematical universes, (or rather, the level I to level IV mathematical multiverses), become increasingly disconnected from empirical reality. (The phrase ``rampant speculation'' comes to mind.)  Now I have used the word multiverse myself~\cite{Wormholes}, but in a very different context and with a very different and much more specific meaning --- when speculating about wormhole physics the various universes in the multiverse are just reasonably large reasonably flat regions of spacetime that are connected to each other via Lorentzian wormholes; and the same (utterly standard) general relativity applies in each universe. 

Experimentalists have an aphorism ``never adjust more than one aspect of your experiment at a time"; theorists should pay heed --- never heap multiple layers of speculation on top of one another. Speculation --- controlled speculation --- is fine; but try to extrapolate only one feature of well-known physics at a time. 
Uncontrolled speculation is a quagmire; a mare's nest;  a necrophiliac deconstruction of the scientific enterprise. 

%-------------------------------------------------------------------------------------------
\section{Discussion} 
%-------------------------------------------------------------------------------------------

\begin{quote}
And I cherish more than anything else the Analogies, my most trustworthy masters. They know all the secrets of Nature, and they ought to be least neglected in Geometry. 
\\
\null\hfill --- Johannes Kepler
\end{quote}

So what message should one take from all this discussion? Overall, I feel that the close relationship between mathematics and physics is not at all surprising --- the reason for the close relationship is in fact utterly prosaic --- ultimately there is a dynamic tension (a dialectic) between the experiments/observations of the natural philosopher and the mathematics then developed to encode the patterns and regularities in the data stream. The natural philosophers and the mathematicians can, and often do, get out of synchronization with each other --- sometimes by centuries --- but overall the most work will go into the mathematics that is the most useful. 

\bigskip

\centerline{---\,\#\#\#\,---}

\bigskip
%-------------------------------------------------------------------------------------------
\section*{Acknowledgments} 
%-------------------------------------------------------------------------------------------

Supported by a Marsden grant, administered by the Royal Society of New Zealand.

\clearpage
%-------------------------------------------------------------------------------------------
%-------------------------------------------------------------------------------------------
%-------------------------------------------------------------------------------------------
\section*{References} 
%-------------------------------------------------------------------------------------------
\renewcommand\refname{}
%-------------------------------------------------------------------------------------------

%-------------------------------------------------------------------------------------------
\clearpage
%------------------------------------------------------------------------------------------
\appendix
%-------------------------------------------------------------------------------------------
\section{Technical end-notes} 
%-------------------------------------------------------------------------------------------
\subsection{Classical uncertainty} 
%-------------------------------------------------------------------------------------------
Consider the purely classical commutators $[t,\partial_t] = - I$, (and similarly $[x,\partial_x] = - I$), with not an $\hbar$ in sight. 
Consider a purely classical signal $s(t)$ and take its Fourier transform
\[
\hat s(\omega) = {1\over\sqrt{2\pi}} \oint s(t) \; e^{-i\omega t} \; \d t.
\]
Now define averages (assume all the relevant integrals converge)
\[
\bar t ={ \oint |s(t)|^2 \; t \; \d t\over  \oint |s(t)|^2 \; \d t} ; 
\qquad 
\bar \omega ={ \oint |\hat s(\omega)|^2 \; \omega \; \d \omega\over  \oint |\hat s(\omega)|^2  \; \d \omega} ;
\]
and variances
\[
\sigma^2_t ={ \oint |s(t)|^2 \; (t-\bar t\,)^2 \; \d t\over  \oint |s(t)|^2 \; \d t} ; 
\qquad 
\sigma^2_\omega ={ \oint |\hat s(\omega)|^2 \;(\omega-\bar\omega)^2 \; \d \omega\over  \oint |\hat s(\omega)|^2 \; \d \omega}.
\]
It is now a \emph{theorem} of mathematics that 
\[
\sigma_t \times \sigma_\omega \geq {1\over2}.
\]
This is the classical time-frequency uncertainty relation. A common interpretation in classical signal processing theory is that the timescale for on-off switching is inversely proportional to the frequency spread in the Fourier transform. This is sometimes phrased as
\[
\hbox{(bit rate)} \lesssim \hbox{(bandwidth)}.
\]
A completely analogous result arises from the $[x,\partial_x]=I$ commutator where, now in terms of position and wave-number, one has
\[
\sigma_x \times \sigma_k \geq {1\over2}.
\]
This is sometimes phrased as
\[
\hbox{(bits per unit length)} \lesssim \hbox{(wave-number spread)}.
\]
It is only once imposes the Einstein relation $E=\hbar\omega$, and the de Broglie relation $p =\hbar k$, that quantum physics is introduced. Specifically, imposing these relations and using what we know about Fourier transforms, we see
\[
E = i \hbar \partial_t;   \qquad\qquad p = -i \hbar \partial_x;
\]
and the usual Heisenberg uncertainty relations now follow
\[
\sigma_t \times \sigma_E \geq {\hbar\over2};  \qquad\qquad \sigma_x \times \sigma_p \geq {\hbar\over2}.
\]
%-------------------------------------------------------------------------------------------
\subsection{Classical barrier penetration} 
%-------------------------------------------------------------------------------------------

Frustrated total internal reflection is a classical barrier penetration effect. It occurs when what would normally be total internal reflection is ``frustrated'' by having only a small gap of low refractive index material separating two regions of high refractive index. 
In this case the ``evanescent wave'' in the gap region (the barrier) allows some of the light to penetrate into the second high refractive index region. 

\noindent
Similar phenomena occur for sound propagation across fluid-fluid-fluid interfaces.

\noindent
The relevant mathematics is formally identical to that required to analyze quantum barrier penetration through a classically forbidden region. 

\noindent 
In short, barrier penetration is primarily a wave effect; it is not (intrinsically) a quantum effect.

\begin{figure}[htbp]
\begin{center}
\includegraphics{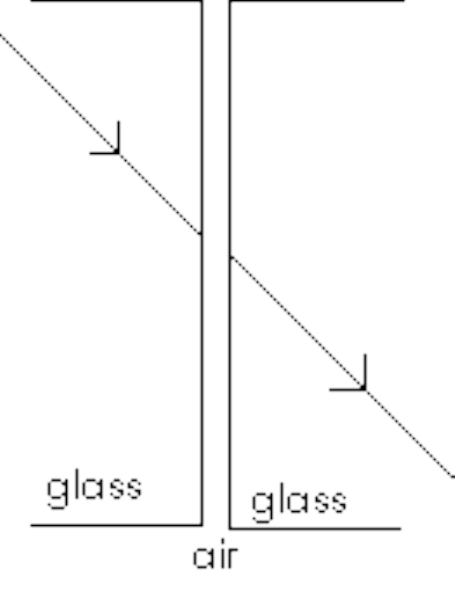}
\caption[scale=0.5]{Frustrated total internal reflection.}
\label{F:ftir}
\end{center}
\end{figure}

\bigskip
\centerline{---\,\#\#\#\,---}

%-------------------------------------------------------------------------------------------
\end{document}